\documentstyle[12pt]{article}

\newcommand{\sect}[1]{\setcounter{equation}{0}\section{#1}}

\def\bseq{\begin{subequation}}  
\def\eseq{\end{subequation}}
\def\bsea{\begin{subeqnarray}}  
\def\esea{\end{subeqnarray}}

\newcommand{\beq}{\begin{equation}}
\newcommand{\eeq}{\end{equation}}
\newcommand{\bea}{\begin{eqnarray}}
\newcommand{\eea}{\end{eqnarray}}
\newcommand {\non}{\nonumber}

\renewcommand{\a}{\alpha}
\renewcommand{\b}{\beta}

\renewcommand{\d}{\delta}
\newcommand{\th}{\theta}
\newcommand{\Th}{\Theta}

\newcommand{\di}{\partial}
\newcommand{\g}{\gamma}
\newcommand{\G}{\Gamma}

\newcommand{\ve}{\varepsilon}

\newcommand{\k}{\kappa}

\renewcommand{\l}{\lambda}

\renewcommand{\o}{\omega}

\newcommand{\calA}{{\cal A}}
\newcommand{\calB}{{\cal B}}

\newcommand{\calE}{{\cal E}}

\def\Mb{\kern 2pt\mathchoice
            {
             \vbox{\hrule width10pt height 0.4pt depth 0pt
                 \kern 1.2pt\hbox{\kern -2pt$\displaystyle M$}}}
            {
                 \vbox{\hrule width10pt height 0.4pt depth 0pt
                 \kern 1.2pt\hbox{\kern -2pt$\textstyle M$}}}
            {
\vbox{\hrule width6pt height 0.4pt depth 0pt
                 \kern 1.0pt\hbox{\kern -2pt$\scriptstyle M$}}}
            {
                 \vbox{\hrule width5pt height 0.4pt depth 0pt
                 \kern 0.8pt\hbox{\kern -2pt$\scriptscriptstyle M$}}}}

\def\Sb{\kern 2pt\mathchoice
            {
                 \vbox{\hrule width6pt height 0.4pt depth 0pt
                 \kern 1.2pt\hbox{\kern -2pt$\displaystyle S$}}}
            {
                 \vbox{\hrule width6pt height 0.4pt depth 0pt
                 \kern 1.2pt\hbox{\kern -2pt$\textstyle S$}}}
            {
                 \vbox{\hrule width3.5pt height 0.4pt depth 0pt
                 \kern 1.0pt\hbox{\kern -2pt$\scriptstyle S$}}}
            {
                 \vbox{\hrule width3pt height 0.4pt depth 0pt
                 \kern 0.8pt\hbox{\kern -2pt$\scriptscriptstyle S$}}}}

\def\Rb{\kern 2pt\mathchoice
            {
                 \vbox{\hrule width5.5pt height 0.4pt depth 0pt
                 \kern 1.2pt\hbox{\kern -2.5pt$\displaystyle R$}}}
            {
                 \vbox{\hrule width5.5pt height 0.4pt depth 0pt
                 \kern 1.2pt\hbox{\kern -2.5pt$\textstyle R$}}}
            {
                 \vbox{\hrule width3.5pt height 0.4pt depth 0pt
                 \kern 1.0pt\hbox{\kern -2.2pt$\scriptstyle R$}}}
            {
                 \vbox{\hrule width3pt height 0.4pt depth 0pt
                 \kern 0.8pt\hbox{\kern -2.2pt$\scriptscriptstyle R$}}}}

  \def\pp{{\mathchoice
          {
              \kern 1pt%
              \raise 1pt
              \vbox{\hrule width5pt height0.4pt depth0pt
                    \kern -2pt
                    \hbox{\kern 2.3pt
                          \vrule width0.4pt height6pt depth0pt
                          }
                    \kern -2pt
                    \hrule width5pt height0.4pt depth0pt}%
                    \kern 1pt
           }
            {
              \kern 1pt%
              \raise 1pt
              \vbox{\hrule width4.3pt height0.4pt depth0pt
                    \kern -1.8pt
                    \hbox{\kern 1.95pt
                          \vrule width0.4pt height5.4pt depth0pt
                          }
                    \kern -1.8pt
                    \hrule width4.3pt height0.4pt depth0pt}%
                    \kern 1pt
            }
            {
              \kern 0.5pt%
              \raise 1pt
              \vbox{\hrule width4.0pt height0.3pt depth0pt
                    \kern -1.9pt  
                    \hbox{\kern 1.85pt
                          \vrule width0.3pt height5.7pt depth0pt
                          }
                    \kern -1.9pt
                    \hrule width4.0pt height0.3pt depth0pt}%
                    \kern 0.5pt
            }
            {
              \kern 0.5pt%
              \raise 1pt
              \vbox{\hrule width3.6pt height0.3pt depth0pt
                    \kern -1.5pt
                    \hbox{\kern 1.65pt
                          \vrule width0.3pt height4.5pt depth0pt
                          }
                    \kern -1.5pt
                    \hrule width3.6pt height0.3pt depth0pt}%
                    \kern 0.5pt
            }
        }}

  \def\mm{{\mathchoice
                       {
                             \kern 1pt
               \raise 1pt    \vbox{\hrule width5pt height0.4pt depth0pt
                                  \kern 2pt
                                  \hrule width5pt height0.4pt depth0pt}
                             \kern 1pt}
                       {
                            \kern 1pt
               \raise 1pt \vbox{\hrule width4.3pt height0.4pt depth0pt
                                  \kern 1.8pt
                                  \hrule width4.3pt height0.4pt depth0pt}
                             \kern 1pt}
                       {
                            \kern 0.5pt
               \raise 1pt
                            \vbox{\hrule width4.0pt height0.3pt depth0pt
                                  \kern 1.9pt
                                  \hrule width4.0pt height0.3pt depth0pt}
                            \kern 1pt}
                       {
                           \kern 0.5pt
             \raise 1pt  \vbox{\hrule width3.6pt height0.3pt depth0pt
                                  \kern 1.5pt
                                  \hrule width3.6pt height0.3pt depth0pt}
                           \kern 0.5pt}
                       }}

\def\pd{{\kern0.5pt
                   + \kern-5.05pt \raise5.8pt\hbox{$\textstyle.$}\kern
0.5pt}}

\def\pmd{{\kern0.5pt
                  \pm \kern-5.05pt \raise6.3pt\hbox{$\textstyle.$}\kern1.5pt}}

\def\md{{\mathchoice
   {
      {{\kern 1pt - \kern-6.2pt \raise5pt\hbox{$\textstyle.$}\kern 1pt}}}
    {
      {{\kern 1pt - \kern-6.2pt \raise5pt\hbox{$\textstyle.$}\kern 1pt}}}
    {
      {\kern0.5pt - \kern-5.05pt \raise3.4pt\hbox{$\textstyle.$}\kern0.5pt}}
    {
      {\kern0.5pt - \kern-5.05pt \raise3.4pt\hbox{$\textstyle.$}\kern0.5pt}}}}

\newcommand{\Del}{\nabla}

\newcommand{\grad}{\nabla}
\newcommand{\sgn}{\mbox{sgn}}

\def\Sc{\scriptstyle}

\newcommand{\reff}[1]{(\ref{#1})}

\newcommand{\shalf}{{\Sc\frac{1}{2}}}
\newcommand{\sihalf}{{\Sc\frac{i}{2}}}
\newcommand{\half}{\frac{1}{2}}

\newcommand{\squart}{{\Sc\frac{1}{4}}}

\renewcommand{\thefootnote}{\fnsymbol{footnote}}

\begin{document}

\newpage
\begin{titlepage}
\begin{flushright}
{hep-th/9809082}\\
{WATPHYS-TH-98/07}\\
{McGill/98-15}\\
\end{flushright}
\vspace{2cm}
\begin{center}
{\bf {\large} Cosmological Supergravity from a Massive Superparticle and
Super Cosmological Black Holes}\\
\vspace{1.5cm}
M.E. Knutt-Wehlau\footnote{knutt@physics.mcgill.ca}\\
\vspace{1mm}
{\em Physics Department, McGill University, Montreal, PQ
CANADA H3A 2T8}\\
\vspace{4mm}
and\\
\vspace{4mm}
R. B. Mann\footnote{mann@avatar.uwaterloo.ca} \\
\vspace{1mm}
{\em Physics Department, University of Waterloo, Waterloo, ON CANADA  N2L 3G1}

\vspace{1.1cm}
{{ABSTRACT}}
\end{center}

\begin{quote}

We describe in superspace a classical theory
of two dimensional $(1,1)$ dilaton supergravity with a
cosmological constant, both with and without coupling to a massive
superparticle.  We give general exact non-trivial superspace solutions
 for the compensator superfield that
describes the supergravity in both cases.
We then use these compensator solutions
to construct models of two-dimensional
supersymmetric cosmological black holes.
\end{quote}

\vfill

\begin{flushleft}
September 1998

\end{flushleft}
\end{titlepage}

\newpage

\renewcommand{\thefootnote}{\arabic{footnote}}
\setcounter{footnote}{0}
\newpage
\pagenumbering{arabic}

\sect{Introduction}

In recent years, the study of D-branes
\cite{deWit} has led to a resurgence
of interest in theories of superparticles in various dimensions, as D0-branes
are massive superparticles. Of particular interest in this context
is the development of exact solutions of supergravity
theories that contain a superparticle.  The list of solutions is quite
meagre: very few exact non-trivial classical
solutions to supergravity theories are known (a list is given in ref.
 \cite{sparticle})
and exact superspace supergravity solutions were non-existent.

This latter situation has recently changed \cite{sparticle}.  By considering
a theory of  a massive superparticle coupled to a version of two-dimensional
$(1,1)$ dilaton supergravity, we obtained  exact
 classical superspace solutions for the superparticle worldline and
the supergravity fields.  Finding non-trivial  solutions (those that cannot
be reduced by infinitesimal  local supersymmetry transformations to purely
bosonic
solutions) to classical supergravity theories is a difficult task, even
in two dimensions.
The (non)-triviality of  a solution can be determined by the method
 given in \cite{aichel}, which requires solving a differential equation for
 an appropriately well-behaved infinitesimal spinor.  However, it is possible
 to sidestep this
procedure by examining classical supergravity problems in superspace
\cite{bible}.  A
 {\em bona fide} superspace supergravity solution -- one which satisfies the
constraints -- has nonzero torsion beyond that of flat superspace.
 The torsion is a
supercovariant quantity, and as such its value remains unchanged under a
gauge transformation. Hence any exact superspace solution with non-zero
torsion must necessarily be non-trivial in this sense.

We take this approach in the present paper. Motivated by the above,
we consider $(1,1)$ supergravity in $(1+1)$ dimensions with a cosmological
 constant,
both with and without coupling to a massive superparticle.
The theory then essentially becomes that of a superLiouville field (a dilatonic
theory of supergravity), with  a  possible superparticle coupling.

For the supergravity part of the theory, we consider as before a supersymmetric
generalization of the $(1+1)$ dimensional ``R=T'' theory, in which
the evolution of the supergravitational fields are determined only by
the supermatter stress-energy (and vice versa).
This ensures that the dilaton field classically decouples from the
evolution of the supergravity/matter system, resulting in a two-dimensional
theory that is most closely aligned with general relativity.
We obtain non-trivial solutions for the supergravity
compensator for the cases of cosmological $(1,1)$ supergravity (with and
without a massive superparticle).  For the case with
the superparticle, we construct a model of a supersymmetric
cosmological black hole.

The outline of our paper is as follows.
In section 2, we review the bosonic solution for
cosmological dilaton gravity (with and without coupling to a massive
particle) and discuss the respective spacetimes.
 In section 3, we outline the compensator solution for
cosmological dilaton $(1,1)$ supergravity,
with no superparticle coupling.
  In section 4 we
describe the superparticle coupled to supergravity in superconformal gauge.
In section 5 we give the solution for the supergravity compensator
 in the presence of this superparticle (details can be found in the
appendix) and discuss the
results. In section 6 we give the supercoordinate
transformations from superconformal to ``super'' Schwarzschild gauge,
 from
which we construct the super black hole models.  In section 7 we present
some concluding remarks.

\sect{Cosmological Dilaton Gravity}

 We start by presenting the bosonic case, that of a massive point particle
interacting with dilaton gravity. This has been done in \cite{tarasov}, but we
briefly review the situation here as it is simpler than the supersymmetric
case, and
illustrates the basic ideas.  We then transform the result
 from Schwarzschild gauge to conformal gauge, and explicitly exhibit
 the requisite
coordinate transformation in order to make contact later on with the
supersymmetric case. We  find it is  necessary
to solve the corresponding supersymmetric theory in superconformal gauge
from the beginning, and transform in the opposite direction to ``super"
Schwarzschild gauge afterwards.

We consider the action for $R=T$ theory, which is
\beq\label{act1}
S=S_G+S_M=\frac{1}{2\kappa}\int d^2x\left[{\sqrt{-g}}({\psi}R+{1\over 2}
(\nabla\psi)^2 )
-\kappa{\cal L}_M\right]
\eeq
where the gravitational coupling $\kappa = 8\pi G$. The action (\ref{act1})
ensures that the dilaton field $\psi$ decouples from the classical
 equations of motion
which, after some manipulation, are
\begin{eqnarray}
R&=&\kappa T_\mu^\mu \label{rt}\\
\frac{1}{2}\left(\grad_\mu\psi\grad_\nu\psi
-\frac{1}{2}g_{\mu\nu}(\grad\psi)^2\right)
-\grad_\mu\grad_\nu\psi+g_{\mu\nu}\grad^2\psi &=& \kappa T_{\mu\nu}
\end{eqnarray}
where $T_{\mu\nu} =\frac{1}{\sqrt{-g}} \frac{\delta {\cal L}_M}
{\delta g^{\mu\nu}}$ is
the stress-energy tensor
and $R_{\mu\nu} = \partial_\lambda\Gamma^\lambda_{\mu\nu}-
\partial_\nu\Gamma^\lambda_{\mu\lambda}-
\Gamma^\lambda_{\mu\sigma}\Gamma^\sigma_{\lambda\nu}
+\Gamma^\lambda_{\lambda\sigma}\Gamma^\sigma_{\mu\nu}$ is our convention
for the Ricci scalar.

Here we take the matter Lagrangian to be that of a cosmological constant,
${\cal L}_M = -\sqrt{-g}\Lambda$, so that $T_{\mu\nu} = \frac{1}{2}g_{\mu\nu}
\Lambda$.
In  conformal coordinates where $ds^2=e^{2\rho}dx^+ dx^-$ equation (\ref{rt})
 becomes
\beq\label{confrt}
\partial_+\partial_-\rho = -\frac{\kappa}{8}\Lambda e^{2\rho}
\eeq
which is Liouville's equation. This equation also follows directly from
the action (\ref{act1}) written in conformal coordinates:
\beq\label{actconf}
S = \frac{1}{2\kappa}\int d^2x \left[-4\psi \partial_+\partial_-\rho
+\partial_-\psi \partial_+\psi + \frac{\kappa}{2}\Lambda e^{2\rho}  \right]
\eeq
This representation will be useful later on in comparing our results to the
supersymmetric case.

Eq. (\ref{confrt}) has the solution
\beq\label{liousol}
\rho = \frac{1}{2}\ln\left[\frac{1}{{\pi G|\Lambda|}}
\frac{f'_+ f'_-}{(f_+ -\sgn(\Lambda)f_-)^2}\right]
\eeq
where $f_{\pm} = f_{\pm}(x^{\pm})$ are two arbitrary functions of the
coordinates
 $(x^+,x^-)$ (where ($x^\pm = (x\pm t)/2$),
and $'$ means derivative with respect to the functional argument.
 The arbitrariness in
the functions $f_\pm$ corresponds to the freedom to choose coordinates.
 Indeed, the metric
associated with (\ref{liousol}) is
\beq\label{lioumet}
ds^2 = \frac{dX^+ dX^-}{{\pi G|\Lambda|}(X^+ -\sgn(\Lambda)X^-)^2}
\eeq
where the coordinates $X^\pm$ have been chosen so that $X^\pm = f_\pm$.

It is always possible to write the metric (\ref{lioumet}) in static coordinates
 by
making the coordinate transformation
\beq\label{stat}
X^\pm = \pm e^{2\sqrt{\pi G|\Lambda|}t} [f(y)]^{\pm 1}
\eeq
yielding
\beq\label{stmet}
ds^2 = \frac{-4\pi G|\Lambda|f^2(y)dt^2+(f'(y))^2dy^2}{\pi G|\Lambda|(f^2
+\sgn(\Lambda))^2}
\eeq
To write the metric in Schwarzschild-type coordinates we  must require
\beq\label{fsch}
\frac{2 f f'}{\sqrt{\pi G|\Lambda|}(f^2+\sgn(\Lambda))^2} = \pm 1
\eeq
which after solving for $f(y)$ gives
\begin{eqnarray}
ds^2 &=& \calA dt^2 + \frac{dy^2}{\calB}
\label{schcosa}
\end{eqnarray}
where
\bea
\calB = -\calA &=&4[(c-\sgn(\Lambda)c^2)\pm\sqrt{\pi
G|\Lambda|}y(1-2\sgn(\Lambda)c)
         -\pi G\Lambda y^2]
\eea
and where $c$ is an arbitrary constant of integration.
 Choosing $c=\sgn(\Lambda)/2$
to eliminate the term linear in $y$ yields
\beq\label{schcos}
ds^2 = -(\sgn(\Lambda) - \frac{\kappa}{2}\Lambda y^2)dt^2
+ \frac{dy^2}{\sgn(\Lambda) - \frac{\kappa}{2}\Lambda y^2}
\eeq
which is the generic form of the metric of a constant curvature spacetime
in Schwarzschild-type coordinates.

For positive $\Lambda$ this is the metric of de Sitter spacetime
\beq\label{2dsit}
ds^2 = -(1 - \frac{\kappa}{2}\Lambda y^2)dt^2
+ \frac{dy^2}{1 - \frac{\kappa}{2}\Lambda y^2}
\eeq
whereas for negative $\Lambda$, we have
\beq\label{2adsitbh}
ds^2 = -(\frac{\kappa}{2}|\Lambda| y^2-1)dt^2
+ \frac{dy^2}{\frac{\kappa}{2}|\Lambda| y^2-1}
\eeq
which is the metric of a $(1+1)$ dimensional anti de Sitter black hole
\cite{tarasov,r5}. Locally this metric is equivalent to
\beq\label{2adsit}
ds^2 = -(\frac{\kappa}{2}|\Lambda| Y^2+1)dT^2
+ \frac{dY^2}{\frac{\kappa}{2}|\Lambda| Y^2+1}
\eeq
which is the usual representation of the anti de Sitter metric in
static coordinates.

We can extend the previous solutions to include the effects of a
single point particle, whose stress energy is given by
\beq\label{ptstress}
T_{\mu\nu} = m\int d\tau \frac{1}{\sqrt{-g}}
   g_{\mu\alpha}g_{\nu\beta}\frac{dz^\alpha}{d\tau}\frac{dz^\beta}{d\tau}
       \delta^{(2)}(x-z(\tau))
\eeq
with $z^\mu(\tau)$ being the worldline of the particle.  Choosing a frame at
rest with respect to the particle, the trace of the stress energy is
\beq\label{btrace}
T_\mu^\mu = -2m e^{-2\rho}\delta(x)
\eeq
in conformal coordinates, $x=0$ being the location of the particle. The
field equations (\ref{rt}) then become
\beq\label{pconfrt}
\partial_+\partial_-\rho = -\frac{\kappa}{8}\Lambda e^{2\rho}
+ \frac{\kappa m}{4}\delta(x)
\eeq
or more simply
\beq\label{lioupart}
\rho''(x) = -\frac{\kappa}{8}\Lambda e^{2\rho} + M\delta(x)
\eeq
with $M= 2\pi Gm$.

Setting $a^2=\frac{\kappa}{8}|\Lambda|$, equation (\ref{lioupart}) has
for $\Lambda >0$ the solution
\beq\label{lpartsol1}
\rho = -\ln\left(\cosh(a|x|+b)\right)
\eeq
where $\tanh(b) = -\frac{M}{2a}$. In Schwarzschild-type coordinates the metric
(\ref{lpartsol1}) becomes
\beq\label{lpart1}
ds^2 = -(-\frac{\kappa}{2}\Lambda y^2+2M|y|+C)dt^2
+ \frac{dy^2}{-\frac{\kappa}{2}\Lambda y^2+2M|y|+C}
\eeq
where $C=1-\frac{M^2}{4a^2}$.  If $\Lambda < 0$, the solution can be
chosen either as
\beq\label{lpartsol2a}
\rho = -\ln\left(\cos(a|x|+b)\right)
\eeq
where $\tan(b) = \frac{M}{2a}$, or as
\beq\label{lpartsol2b}
\rho = -\ln\left(\sinh(a|x|+b)\right)
\eeq
where $\coth(b)=-\frac{M}{2a}$.  In Schwarzschild-type coordinates these
become respectively
\beq\label{lpart2a}
ds^2 = -(\frac{\kappa}{2}|\Lambda| y^2 + 2M|y|+C)dt^2
+ \frac{dy^2}{\frac{\kappa}{2}|\Lambda| y^2 + 2M|y|+C}
\eeq
where $C=1+\frac{M^2}{4a^2}$ and
\beq\label{lpart2b}
ds^2 = -(\frac{\kappa}{2}|\Lambda| y^2 + 2M|y|-C)dt^2
+ \frac{dy^2}{\frac{\kappa}{2}|\Lambda| y^2 + 2M|y|-C}
\eeq
where $C=1-\frac{M^2}{4a^2}$.

The solutions (\ref{lpart1}), (\ref{lpart2a}) and (\ref{lpart2b}) have been
discussed previously in ref. \cite{tarasov}. For positive $M$ and $C$, the
metric (\ref{lpart1}) describes a point
mass in de Sitter spacetime.  If $M>0$ but $C<0$, then (\ref{lpart1}) describes
the $(1+1)$-dimensional analogue of Schwarzschild de Sitter spacetime: there is
both a cosmological and an event horizon.  The metric (\ref{lpart2a}) is
that of a point mass in anti de Sitter space, and the metric in
(\ref{lpart2b}) is the $(1+1)$-dimensional analogue of a Schwarzschild
anti de Sitter black hole provided $C>0$.

\sect{Cosmological $(1,1)$ Dilaton Supergravity}

    We consider next in superspace a
 theory of $(1,1)$ dilaton supergravity in two dimensions
with a cosmological constant, $L$.  We use light-cone coordinates
$(x^\pp, x^\mm)= \half (x^1 \pm x^0)$ and $(\th^+, \th^-)$. The action  is
given by
\beq\label{3p1}
I_C = -\frac{2}{\k}\int d^2xd^2 \th E^{-1}(\Del_+ \Phi \Del_- \Phi
      + \Phi R -4L)
\eeq
where $\Phi$ is the dilaton superfield,
 $R$ is the scalar supercurvature, $E = {\rm sdet} {E_A}^M$, the
superdeterminant
of the vielbein, and $\Del_\pm$ is the supercovariant derivative, defined
below.
 (As in \cite{sparticle}, we choose this particular action since the dilaton
decouples from the evolution of the matter system in this case,
giving a theory  that most closely resembles a supersymmetric analogue of
two-dimensional general relativity.)

The solution to the constraints on the covariant derivatives
is simplest in conformal gauge.
 The covariant derivatives,
 $\Del_A = {E_A}^M D_M + \o_A M$ where $\o_A$ is the spin connection,
 are expanded with respect to the standard flat
supersymmetry covariant derivatives, $D_A = (D_+, D_-) = (\di_+ + i \th^+
\di_\pp,
\di_- + i \th^- \di_\mm)$. However when we consider
the superparticle coupling,
the natural description will be in terms of forms, and so we choose as a basis
the ordinary derivatives $\di_M = (\di_m, \di_\mu)$, where
 we write $\Del_A = {{\calE}_A}^M \di_M + \o_A M$.
We solve the constraints
in conformal gauge in terms of the $D$'s and change
to the other basis afterwards.
 The (1,1) supergravity constraints \cite{jim, rocek} are
\bea
\{\Del_+, \Del_+\} = 2i \Del_\pp  &,& \{\Del_-, \Del_-\} = 2i \Del_\mm \non \\
\{\Del_+, \Del_-\} &=& RM \non \\
{T_{+\pp}}^A &=& {T_{-\mm}}^A = 0  \label{cons}
\eea
where the covariant derivatives are determined in conformal
gauge in terms of the compensator superfield $S$, as
\bea
\Del_+ = e^S [D_+ + 2 (D_+S)M] &,& \Del_- = e^S [D_- - 2 (D_-S)M] \non\\
R &=& 4e^{2S}D_- D_+ S \label{R}
\eea
{}from which we can read off the
elements of
${E_A}^M$ and compute $E^{-1} = e^{-2S}$.

  In the preferred basis we calculate the elements of
${{\calE}_A}^M$, and
inverting this matrix we obtain
\bea
{{\calE}_M}^A = \left[
\begin{array}{cccc}
e^{-2S} & 0 & 2ie^{-S}D_+S & 0 \\
0 & e^{-2S} & 0 & 2ie^{-S}D_-S \\
-ie^{-2S}\th^+ & 0 & e^{-S}(1-2(D_+S) \th^+) & 0 \\
0 & -i e^{-2S} \th^- & 0 & e^{-S}(1-2(D_-S) \th^-)
\end{array}
\right]
\eea
Therefore in conformal gauge, the cosmological
dilaton supergravity action
becomes
\beq
I_C = -\frac{2}{\k}\int d^2x d^2\th (D_+ \Phi D_- \Phi + 4 \Phi D_- D_+ S
       - 4 e^{-2S} L)
\label{ccosmo}
\eeq
The equations of motion are
\bea
 \Phi  &=& - 2 S\label{eqphi} \\
 D_- D_+ S &=&  e^{-2S} L
\label{mateq}
\eea
Thus the dilaton is expressible in terms of the compensator $S$,
which satisfies the usual  superLiouville equation.

In order to recover the component action of \reff{actconf}, we identify
the components of the superfields by the theta expansion (dropping the
fermionic fields),
\bea
S &=& -\frac{1}{2}\rho + \sigma \th^+\th^- \label{Scomp}\\
\Phi &=& -\frac{1}{2}\psi + \varphi \th^+\th^- \label{Phicomp}
\eea
which yields
\beq\label{nactconf}
I_C = \frac{2}{\kappa}\int d^2x \left[-\psi \partial_\pp\partial_\mm\rho
+\frac{1}{4}\partial_\mm\psi \partial_\pp\psi -\varphi^2 - 4\sigma\varphi
-8\sigma L e^{\rho}  \right]
\eeq
from the superspace action (\ref{ccosmo}).  Upon elimination of the auxiliary
fields $\varphi,\sigma$ via their equations of motion,
 (\ref{nactconf})
becomes
\beq
I_C = \frac{1}{2\kappa}\int d^2x \left[-4\psi \partial_\pp\partial_\mm\rho
+\partial_\mm\psi \partial_\pp\psi - 16 L^2 e^{2\rho}  \right]
\eeq
which is equivalent to (\ref{actconf}) provided
$\Lambda = -\frac{32}{\k}L^2$. Hence only the action for anti de Sitter
spacetimes is recovered.
Alternatively, inserting the superfield expansions (\ref{Scomp},\ref{Phicomp})
into eqs. (\ref{eqphi}) and (\ref{mateq}) yields after some manipulation
\beq\label{sconfrt}
\partial_\pp\partial_\mm\rho = 4 L^2 e^{2\rho}
 = -\frac{\kappa}{8}\Lambda e^{2\rho}
\eeq
which is the Liouville equation (\ref{confrt}).

  The  form of the solution to the superLiouville equation (\ref{mateq})
that we find most useful is given by \cite{arvis}.
 We write the result so that it corresponds more closely with the
bosonic result in \cite{rliou} and find
\beq\label{sliousol}
S = -\frac{1}{2} \ln \left[ \frac{-i}{2L}
          \frac{(D_+F_+) (D_- F_-)} {({\cal F}_\pp -
 {\cal F}_\mm - iF_+F_-)}\right]
\eeq
where $F_+ = F_+(x^\pp, \th^+)$ and $F_- = F_-(x^\mm, \th^-)$ are spinor
superfields, with $D_- F_+ = D_+ F_- = 0$, and ${\cal F}_{\pp/\mm}$
satisfying
\bea
D_+ {\cal F}_\pp &=& i F_+ D_+F_+ \non \\
D_- {\cal F}_\mm &=& i F_- D_-F_-
\eea
These equations are solved by
\bea
{\cal F}_\pp &=& f_\pp \pm i \th^+ \l^+\sqrt{\di_\pp f_\pp} \non \\
F_+ &=& \pm \sqrt{\di_\pp f_\pp}\left[ 1 + \sihalf \l^+
               \frac{\di_\pp\l^+}{\di_\pp f_\pp}\right] \th^+ + \l^+ \non \\
{\cal F}_\mm &=& f_\mm \pm i \th^- \l^-\sqrt{\di_\mm f_\mm} \label{410} \\
F_- &=& \pm \sqrt{\di_\mm f_\mm}\left[ 1 + \sihalf \l^-
               \frac{\di_\mm\l^-}{\di_\mm f_\mm}\right] \th^- + \l^- \non
\eea
where $f_\pp = f_\pp(x^\pp), \l^+ = \l^+(x^\pp),
 f_\mm = f_\mm(x^\mm)$ and $\l^- = \l^-(x^\mm)$.
It is straightforward to show using eqs. (\ref{410}) and (\ref{Scomp})
that the leading term in the superfield
expansion of (\ref{sliousol}) is given by  (\ref{liousol}).

Since the functions $\l^\pm$ and $f_{\pp/\mm}$ are arbitrary, we
consider the following special case:  we simplify the expression
 (\ref{sliousol}) by choosing coordinates so that
\beq\label{schoice}
D_+F_+ = 1 = -iD_-F_-
\eeq
Upon insertion of (\ref{410}) this yields $f_{\pp/\mm}=\pm x^{\pp/\mm}$,
$\l^+ = \l^+_0$ and $\l^- = i\l^-_0$, where $\l^+_0$ and $\l^-_0$ are
anticommuting constants.  This gives
\beq\label{sliousimp}
S = \frac{1}{2} \ln \left[ 2L
          (x^\pp + x^\mm + i(\th^+\l^+_0 +\th^-\l^-_0)
 +(\th^+ + \l^+_0)(\th^- + \l^-_0))  \right]
\eeq
for the compensator.  The significance of this expression is perhaps more
easily understood if we examine it in terms of component fields.  We follow
the procedure given in \cite{sparticle,rocek} in which we choose a
Wess-Zumino gauge such that the superconformal
gauge ($E_\pm = e^S D_\pm$) is compatible with the ordinary $x$-space conformal
gauge ($e_a^m = e^{\rho} \delta_a^m$ in appropriate coordinates).
Using these results, we write $E_\a$ in component
conformal gauge as
\bea
E_\pm &=& e^S D_\pm \non \\
 &=& [ e^{-1/4} + \sihalf(\th^+ {\psi_\pp}^+ - \th^- {\psi_\mm}^-) +
\squart \th^+ \th^-
            e^{1/4}(iA + {\psi_\mm}^- {\psi_\pp}^+)] D_\pm
\eea
and then Taylor-expand $e^S$ so that we can
 identify the gravitino (${\psi_\pp}^+, {\psi_\mm}^-$) and the
 component auxiliary field of the supergravity multiplet, $A$,
 by comparing powers
of $\th$.

Carrying out the expansion, we find
\bea
e^{-\rho/2} &=& \sqrt{2L(x^\pp+x^\mm)}
\left(1+\frac{\l^+_0\l^-_0}{2(x^\pp+x^\mm)}\right)
         \non \\
{\psi_\pp}^+ &=&  \sqrt{\frac{2L}{(x^\pp+x^\mm)}}
\left( \l^+_0 + \l^-_0 \right)
                      \non \\
{\psi_\mm}^- &=&  \sqrt{\frac{2L}{(x^\pp+x^\mm)}}
\left( \l^+_0 - \l^-_0 \right) \\
A  &=&  -4iL \non
\label{slcomponents}
\eea
for the component fields.  Note that the presence of
 both gravitini fields is needed
in order for the correction to the conformal
 factor $\rho$ to be non-zero.  The metric in Schwarzschild coordinates
can be found by
 transforming to static coordinates
\beq\label{slstat}
x^{\pp/\mm} = \pm e^{4Lt} [f(y)]^{\pm 1}
\eeq
as in the bosonic case, and transforming the component fields appropriately.
 For $f(y)$
given by
\beq\label{sladS}
f(y) = \frac{8Ly+1}{8Ly-1}
\eeq
which is the solution to (\ref{fsch}) with $\Lambda<0$,
we obtain the supersymmetric
analogue of the $(1+1)$-dimensional black hole (\ref{2adsitbh}).

\sect{Superparticle Coupled to Supergravity in Superconformal Gauge}

 We consider now the general action for a massive superparticle coupled to
$(1,1)$ ``R=T'' dilaton supergravity in two dimensions.  We reproduce here
the relevant section of \cite{sparticle}, as the necessary steps for obtaining
 the equation of motion for the compensator
superfield are the same, in this case.
The action is \cite{sparticle}
\bea
I_P &=& m \int d\tau \left[ g^{-1} \dot{z}^M {{\calE}_M}^\pp  \dot{z}^N
{{\calE}_N}^\mm  +
 \dot{z}^M {{\calE}_M}^A \G_A + \frac{g}{4}  \right]
\eea
where  $\G$ is a general gauge
superfield for the Wess-Zumino type term, and $g$ is the einbein on the
worldline of the superparticle. With
 $\widehat{\Del}_A = \Del_A + \G_A$, including now the gauge field,  we have
\beq
[\widehat{\Del}_A, \widehat{\Del}_B\} = {T_{AB}}^C \widehat{\Del}_C + R_{AB} M
+ F_{AB}
\eeq
which defines the torsions, curvatures and gauge field strengths, respectively,
where $\di_{[M} {{\calE}_{N)}}^A = {T_{NM}}^A + {\o_{[MN)}}^A$, and
$F_{AB} = \Del_{[A} \G_{B)} - {T_{AB}}^C \G_C$,
with $\{\G_A, \G_B ] = 0$. The constraints on $\G$ are
\bea
\G_\pp = - i \Del_+ \G_+  &,& \G_\mm = - i \Del_- \G_- \non \\
F_{+-}=F_{-+} &=& \Del_+ \G_- +\Del_- \G_+ = i  \label{Gammacons}
\eea
and all other $F$'s are zero,
consistent with the Bianchi identities \cite{jim}.
The action is $\k$-invariant
 under transformations of the coordinates in curved superspace provided
 that the supergravity constraints and those on $\G$ are satisfied.

  The action for the superparticle in superconformal gauge is
\beq
I_P =  m \int d^4z \int d\tau \left[ g^{-1} \dot{{z_0}}^M {{\calE}_M}^\pp
\dot{{z_0}}^N {{\calE}_N}^\mm
  + \dot{{z_0}}^M  {{\calE}_M}^A \G_A + \frac{g}{4} \right] \d (z -
{z_0}(\tau))
\eeq
where $z = (x, \th)$ are the coordinates of the superspace, and ${z_0}(\tau) =
( {x_0}(\tau),
{\th_0}(\tau))$ are the coordinates of the superparticle.
Defining $\{\widehat{\Del}_+, \widehat{\Del}_+\} \equiv 2i
\widehat{\Del}_
\pp$, and similarly
for $\mm$,  we find the constraints on $\G$ in conformal gauge to be
\bea
\G_\pp &=& - ie^S[D_+ \G_+ + (D_+S)\G_+]  \non \\
\G_\mm &=& - ie^S[D_- \G_- + (D_-S)\G_-] \label{vecGamcons} \\
0 &=& D_+(e^{-S}\G_-) + D_-(e^{-S}\G_+) - i e^{-2S}
\eea
Substituting for $\calE$ and $\G$ we obtain
\bea
I_P &=&  m \int d^4z \int d\tau \left\{ g^{-1} e^{-4S}(\dot{{x_0}}^\pp + i
{\th_0}^+ \dot{{\th_0}}^+)
            (\dot{{x_0}}^\mm + i {\th_0}^- \dot{{\th_0}}^-) \right.\non \\
            &+& i e^{-S}(\dot{{x_0}}^\pp + i {\th_0}^+ \dot{{\th_0}}^+)[(D_+S)
\G_+ - D_+ \G_+] \non \\
            &+& i e^{-S}(\dot{{x_0}}^\mm + i {\th_0}^- \dot{{\th_0}}^-)[(D_-S)
\G_- - D_- \G_-] \non\\
            &+&\left. e^{-S}(\dot{{\th_0}}^+ \G_+  + \dot{{\th_0}}^- \G_-)  +
\frac{g}{4}
       \right\}\d (z-{z_0}(\tau))
\label{part}
\eea

  It is convenient to define $G_\a = e^S \G_\a$ and include \reff{Gammacons}
in the supergravity action by means of a lagrange multiplier, $\l$.
We obtain
\bea
I_P &=&  m \int d^4z \int d\tau \left[ g^{-1} e^{-4S}(\dot{{x_0}}^\pp + i
{\th_0}^+ \dot{{\th_0}}^+)
            (\dot{{x_0}}^\mm + i {\th_0}^- \dot{{\th_0}}^-) \right. \non \\
            &+& i (\dot{{x_0}}^\pp + i {\th_0}^+ \dot{{\th_0}}^+)D_+G_+
                   + i (\dot{{x_0}}^\mm + i {\th_0}^- \dot{{\th_0}}^-)D_-G_-
\non\\
            &+& \left. \dot{{\th_0}}^+ G_+  + \dot{{\th_0}}^- G_-  +
\frac{g}{4} \right] \d (z-{z_0}(\tau))
 \label{act}
\eea
and
\bea
I_C &=& -\frac{2}{\k}\int d^2x d^2\th [D_+ \Phi D_- \Phi + 4 \Phi D_- D_+ S
   - 4 e^{-2S} L \non \\
  &+& \k \l e^{-2S} (D_+G_- + D_-G_+ - i e^{-2S})]
 \label{condil}
\eea

  We now perform a change of variables in the superparticle action, by first
explicitly
writing it in terms of ${x_0}^0$ and ${x_0}^1$, and then making the gauge
choice ${x_0}^0 = \tau$ (static
gauge)
so that $\frac{\dot{{x_0}}^1}{\dot{{x_0}}^0} = \frac{d{x_0}^1}{d{x_0}^0} \equiv
\dot{{x_0}}$.  We also relabel
$z^M = (x^0, x^1, \th^\mu) = (t, x, \th^\mu)$, so \reff{act} becomes
\bea
I_P &=&
m \int dt dx d^2 \th \int d{x_0}^0 \left\{ g^{-1} e^{-4S}
\left[\shalf
(1+\dot{{x_0}}) +
            i {\th_0}^+ \dot{{\th_0}}^+ \right]
             \left[\shalf ( 1 -\dot{{x_0}}) + i {\th_0}^- \dot{{\th_0}}^-
\right] \right. \non \\
            &+& i \left[ \shalf ( 1 +\dot{{x_0}}) + i {\th_0}^+ \dot{{\th_0}}^+
\right]D_+G_+
                + i \left[ \shalf ( 1 -\dot{{x_0}}) + i {\th_0}^-
\dot{{\th_0}}^-\right] D_-G_-  \non\\
            &+& \left. \dot{{\th_0}}^+ G_+  + \dot{{\th_0}}^- G_-  +
\frac{g}{4} \right\}
            \d (t-{x_0}^0) \d (x-{x_0}({x_0}^0)) \d (\th^+ -
{\th_0}^+({x_0}^0)) \d (\th^- - {\th_0}^-({x_0}^0))\non \\
\eea
and
doing the ${x_0}^0$ integration gives
\bea
I_P &=&  m \int dt dx d^2 \th \left\{ g^{-1} e^{-4S} \left[\shalf
(1+\dot{{x_0}}) + i {\th_0}^+
            \dot{{\th_0}}^+ \right]
            \left[\shalf ( 1 -\dot{{x_0}}) + i {\th_0}^- \dot{{\th_0}}^-\right]
\right. \non \\
            &+& i \left[ \shalf ( 1 +\dot{{x_0}}) + i {\th_0}^+ \dot{{\th_0}}^+
\right] D_+G_+
                 + i \left[ \shalf ( 1 -\dot{{x_0}}) + i {\th_0}^-
\dot{{\th_0}}^- \right] D_-G_-  \non\\
&+& \left. \dot{{\th_0}}^+ G_+  + \dot{{\th_0}}^- G_- + \frac{g}{4}
\right\} \d (x-{x_0}(t))
            \d (\th^+ - {\th_0}^+(t)) \d (\th^- - {\th_0}^-(t)) \label{part2}
\eea

  {}From the sum of \reff{condil} and \reff{part2},
 we obtain for the equation of motion for $S$
\bea
 D_- D_+ S(z)  &-&  e^{-2S} L  \non \\
   &=& \frac{\k m}{4} \int dt' \left\{ g^{-1} e^{-4S} \left[\shalf
(1+\dot{{x_0}})
         + i {\th_0}^+ \dot{{\th_0}}^+ \right]
            \left[ \shalf ( 1 -\dot{{x_0}}) + i {\th_0}^-
\dot{{\th_0}}^-\right] \right\} \d^4 (z - {z_0}(t'))
      \non\\
&=& \frac{\k m}{8} \sqrt{\pi^2} e^{-2S} \d (x-{x_0}(t)) \d (\th^+ -
{\th_0}^+(t))
    \d(\th^- - {\th_0}^-(t))
\label{comp}
\eea
where $ \sqrt{\pi^2} = \shalf \sqrt{1-\dot{{x_0}}^2}$
for a free particle.

In solving the above equation for $S$, we strictly speaking must
 include the constraint
 on  the G's as well, in order to have a real solution.  Unlike  the previous
superparticle case of \cite{sparticle}, where the superparticle was allowed to
move freely and $S$ vanished on the worldline of the superparticle,
 this case is more complicated.
We now have a non-vanishing compensator, and as a consequence, the constraint
on the G's is now non-trivial. In the next section, we look only for solutions
 with the superparticle held fixed.

\sect{Solution for Compensator}

 To solve for the compensator $S$ that describes the supergravity generated
by a superparticle in the presence of a cosmological constant, we consider
the superparticle to be stationary and fixed at the origin ($x_0 = \th_0 = 0$)
to begin with.
We find the solution for a superparticle located at a general
 point $(x_0, \th_0)$ by a supersymmetry transformation of this simpler result.
   In this case,
\reff{comp} becomes
\beq
e^{2S} D_- D_+ S(z)  - L
= K \d (x) \d (\th^+) \d(\th^-)
\label{sliou}
\eeq
 where $K = \frac{\k m}{16}$,  and $\sqrt{\pi^2} = \shalf$ for $\dot{{x_0}}=
0$.
 We rewrite the equation in terms of $T = e^{2S}$
\beq
T D_+ D_- T - D_+T D_-T = -2T(L + K \delta(x) \delta^{(2)}(\th))\label{eqnT}
\eeq
where we have set $x_0=\th_0 = 0$.
 We solve for $T$ by doing a $\th$-expansion, $T(x, \th) =
 A(x) + B_+(x) \th^+ + C_-(x)
\th^- + D(x) \th^+ \th^-$, substituting into \reff{eqnT}, and matching powers
of
$\th$. (Note that $\delta^{(2)}(\th) = \th^+\th^-$.)
   We find the following four equations as a result:
\bea
&\th^0: & - AD - B_+C_- = -2 L A  \label{1}\\
&\th^+:& iA{C'}_- - 2DB_+ - i A'C_- = -2L B_+  \label{2}\\
&\th^-:& iA{B'}_+ + 2DC_- - i B_+ A' = + 2 L C_-  \label{3}\\
&\th^+\th^-:& A A'' - 2i B_+ {B'}_+ - 2i C_- {C'}_- +2D^2 -{A'}^2 = 2 L D
    +2KA \delta(x) \label{4}
\label{A-D}
\eea
where the prime denotes derivative with respect
 to $x$.

 The detailed solution to these four equations for the component fields of $T$
is given in the appendix.
 One solution is found to be
 $T(x, \th) = A(x) + B_+(x) \th^+ + C_-(x) \th^- + D(x) \th^+ \th^-$, where
\bea
A &=& 2Lc_0\cos(\frac{|x|-c_1}{c_0}) +\g \left[ 2L{c_0}^2
           + c_3|x| \sin(\frac{|x|-c_1}{c_0})
              + c_3c_0\cos(\frac{|x|-c_1}{c_0})\right] \non \\
\\
B_+ &=& 2Lc_0[ i\a + \b \ve(x)\sin(\frac{|x|-c_1}{c_0})]\\
C_- &=& 2Lc_0[\a \ve(x)\sin(\frac{|x|-c_1}{c_0}) - i \b] \\
D &=&  2L[1 - \g c_0\cos(\frac{|x|-c_1}{c_0})]
\eea
where $\a$ and $\b$ are arbitrary Grassmann constants of integration ($\g = \a
\b$)
and the $c_i$'$s$ are ordinary integration constants.
 This can be put in closed form, which we  understand by a Taylor series
expansion, as
\beq
T(x, \th) = 2L(\th^+ - i \b c_0)(\th^- -i \a c_0)
               +  2Lc_0(1+\frac{ c_3 \g}{2L})
              \cos \left[(\frac{|x| - c_1}{c_0}) - \b\th^+ - \a \th^- - \frac{
                       c_3 \g}{2Lc_0} |x| \right]
 \label{trigT}
\eeq

The second solution for the superfield $T$ is
 $T(x, \th) = A(x) + B_+(x) \th^+ + C_-(x) \th^- + D(x) \th^+ \th^-$, where
\bea
A &=& 2L c_0\sinh(\frac{|x|-c_1}{c_0})  -\g \left[ 2L{c_0}^2
           + c_3 |x|\cosh(\frac{|x|-c_1}{c_0})
         - c_3c_0\sinh(\frac{|x|-c_1}{c_0}) \right] \non\\
     \\
B_+ &=&  2Lc_0[ i\a \ve(x)\cosh(\frac{|x|-c_1}{c_0}) -\b] \\
C_- &=&  -2Lc_0[\a +i\b \ve(x) \cosh(\frac{|x|-c_1}{c_0}) ]\\
D &=&  2L[ 1- \g c_0~\sinh(\frac{|x|-c_1}{c_0})]
\eea
 In closed form, this is
\beq
T(x, \th) = 2 L(\th^+ - \a c_0)(\th^- + \b c_0)
              + 2Lc_0 (1 + \frac{c_3 \g}{2L})
      \sinh \left[(\frac{|x| - c_1}{c_0}) + i\a\th^+
    -i\b \th^- -\frac{c_3 \g}{2Lc_0}|x| \right]  \label{hypT}
\eeq

  As mentioned at the beginning of this section, general solutions for
the case when
the superparticle is located at an arbitrary point $(x_0, \th_0)$
can be obtained by a supersymmetry
transformation of the above results.
Performing a finite supersymmetry transformation on \reff{trigT}
 with the usual
generators $P$ and $Q$ gives the new $T$
\bea
&&e^{i(x_0^\pp P_\pp +x_0^\mm P_\mm + \th_0^+ Q_+ + \th_0^- Q_-)} T(x, \th)
e^{-i(x_0^\pp P_\pp +x_0^\mm P_\mm + \th_0^+ Q_+ + \th_0^- Q_-)} \non \\
&=& T(x-x_0-i(\th^+ \th_0^+ + \th^- \th_0^-), \th -\th_0) \non \\
&=& T(x', \th')
\eea
where $T(x', \th') =  A(x') + B_+(x') {\th^+}' + C_-(x') {\th^-}' + D(x')
{\th^+}'{\th^-}'$.
Defining $X \equiv x-x_0-i(\th^+ \th_0^+ + \th^- \th_0^-)$,
 we can write the transformed solution as
\bea
T(x, \th) &=& 2L(\th^+ -\th_0^+ - i \b c_0)(\th^- -\th_0^- -i \a c_0) \\
  &+&  2Lc_0(1+\frac{c_3 \g}{2L})
              \cos \left[(\frac{|x| - c_1}{c_0}) - \b(\th^+ - \th_0^+)
                     - \a (\th^- - \th_0^-) - \frac{
                       c_3 \g}{2Lc_0} |X|\right] \non
\eea
where
\bea
|X| &\equiv& |x-x_0 - i(\th^+ \th_0^+ + \th^- \th_0^-)| \\
 &=&  |x-{x_0}| - i(\th^+ {\th_0}^+ + \th^- {\th_0}^-)[\Th
(x-{x_0}) -\Th ({x_0}-x)]
     + 2 \th^+ \th^- {\th_0}^+ {\th_0}^- \d (x-{x_0}) \non
\eea
The shifted
hyperbolic solution is obtained in a similar manner.

 The solution depends upon the
parameters $x_0$, $\theta_0$, $\alpha$ and $\beta$.  In particular, the
gravitino field is found  to be
\bea
{\psi_\pp}^+ &=& - 4\th_0^+ \frac{d}{dx}\sqrt{A - C_- \th_0^-} +
            2i\frac{B_+ + D \th_0^-}{\sqrt{A - C_- \th_0^-}}     \non \\
{\psi_\mm}^- &=& 4 \th_0^- \frac{d}{dx}\sqrt{A - B_+\th_0^+}  -
            2i\frac{C_- - D\th_0^+}{\sqrt{A - B_+\th_0^+}}
\eea
where $A-D$  are given by \reff{A-D} with $x$ replaced by $x-x_0$. The
parameters
 $\alpha$ and $\beta$ are arbitrary constants of integration. We note that
choosing them to be  zero makes the gravitino field proportional to $\theta_0$.
 This is
not surprising since  the  superparticle
component supercurrent - a source for the gravitino field -
 is itself proportional to $\th_0$. For general $\alpha$, $\beta$, it is
 possible to
choose $\th_0$ so as to make one of the gravitino components vanish, but
except for the case $\alpha = \beta = \theta_0 =0$, the gravitino field is
 nonvanishing.

  The non-triviality of the solutions can be seen  by calculating
the  torsion ${T_{\pm,\mp \mp}}^\a$, which is
proportional to $R$ and obviously not zero,
 or the torsion ${T_{\pp,\mm}}^\a$, which is proportional
to $\Del_\pm R$ \cite{rocek}. The latter can be shown to
be non-zero, even when $\a =\b = \th_0 = 0$.  Hence our
result has the requisite non-zero torsion, and is  therefore not
 just a gauge transform
of a purely gravitational solution in  dilaton gravity.

\sect{Schwarzschild Gauge}

   So far we have used the superconformal coordinates $z=(x, \th)$
to parametrize the superspace.  To facilitate comparison with the
results of \cite{tarasov}, we transform
now to superspace coordinates $w=(u, \l)$ that
correspond to Schwarzschild gauge, in which the vielbein of the bosonic
subspace takes the form
\bea\label{2dzweibos}
{e_m}^a = \left[
\begin{array}{cc}
\sqrt{\a} & 0  \\
0 & {\sqrt{\a}}^{-1}
\end{array}
\right]
\eea
For simplicity we shall set $c_3=0$ throughout.

By analogy with the bosonic case,
the supercoordinate transformation that takes us from
 $z=(x, \th)$ to $w=(u, \l)$ for $T$ given by \reff{trigT} is
\bea\label{slpart1}
2L(|x|-c_1) &=& \tan^{-1}[ 4L(|u| + u_0)] \\
\l^+ &=&  \frac{\th^+ - \frac{i\b}{2L}}{\frac{1}{4L}
 \tan[2L(|x| -c_1)]+ \a \th^+ + \b \th^-} \\
\l^- &=&  \frac{\th^- -  \frac{i\a}{2L}}{\frac{1}{4L}
 \tan[2L(|x| -c_1)] + \a \th^+ + \b \th^-}
\eea
and the transformation from superconformal to Schwarzschild
coordinates for $T$ of
\reff{hypT} is
\bea\label{slpart2}
2L(|x|-c_1) &=& \coth^{-1}[ 4L(|u| + u_0)] \\
\l^+ &=&  \frac{\th^+ - \frac{\a}{2L}}{\frac{1}{4L}
 \coth[2L(|x| -c_1)] + i \b \th^+ -i \a \th^-} \\
\l^- &=&  \frac{\th^- + \frac{\b}{2L}}{\frac{1}{4L}
 \coth[2L(|x| -c_1)] + i \b \th^+ - i\a \th^-}  \qquad .
\eea
The transformations (\ref{slpart1}) and (\ref{slpart2}) are those which bring
the
bosonic part of the vielbein into a form which reproduces the
supersymmetric extension of the metrics (\ref{lpart2a}) and (\ref{lpart2b})
respectively.  The latter case corresponds to that of a super
black hole.  When applied to remaining part of the vielbein,
these transformations yield expressions for the gravitini and
the auxiliary fields once a component expansion is carried out.  We
shall not reproduce these expressions here.

\section{Summary}

We have obtained several new exact solutions in $(1+1)$ dimensional
supergravity
in superspace.  By coupling a superdilaton to supergravity as in (\ref{3p1}),
we are able to construct a theory in which the stress-energy of supermatter
generates supercurvature, without any influence (classically)
 from the superdilaton
field.  If the supermatter is chosen to be that associated with a
cosmological constant,
we find that the cosmological constant is necessarily negative.
 Two exact solutions
to the field equations are then obtained. One is a supersymmetric version
of anti de Sitter spacetime. The other is a supersymmetric version of the
anti de Sitter black hole discussed in refs. \cite{tarasov,r5}.

Once a superparticle is coupled to the system, the situation changes. The
constraints on the gauge superfield $\Gamma$ are no longer trivial (as compared
to the case with zero cosmological constant), and so solving for the
compensator
becomes somewhat more complicated.  Two exact solutions are again obtained,
which are supersymmetric analogues of their bosonic counterparts
(\ref{lpartsol2a}) and (\ref{lpartsol2b}). The latter solution corresponds to
a massive supersymmetric black hole in anti de Sitter spacetime;  it can be
considered the supersymmetric $(1+1)$ dimensional analogue of Schwarzschild
anti de Sitter spacetime.

We have been able to obtain exact superspace solutions
for the cases we have considered. The role these solutions play in
two-dimensional quantum supergravity remains to be studied.

\noindent {\bf Acknowledgments}

This research was
supported in part by the Ontario-Qu\'{e}bec Projects of Exchange at the
University Level, NSERC of Canada,
and an NSERC Postdoctoral Fellowship.

\vspace{0.5cm}
\appendix
\sect {\bf Appendix}

\setcounter{equation}{0}

We solve the four equations in the text for the component fields of $T$
 ($A, B_+, C_-$ and $D$).   In doing so, we find it convenient to rewrite
\reff{1}--\reff{4} as follows:
\bea
D -2L &=& -\frac{B_+ C_-}{A} \label{1'}\\
\left(\frac{C_-}{A}\right)' &=& -2i(D -L) \frac{B_+}{A^2} \label{2'}\\
\left(\frac{B_+}{A}\right)' &=& 2i(D -L) \frac{C_-}{A^2} \label{3'} \\
\frac{A''}{A} - \frac{{A'}^2}{A^2} - 2i  \frac{B_+}{A} \frac{B'_+}{A}
   - 2i \frac{C_-}{A} \frac{C'_-}{A} &=&
   -2\frac{D^2}{A^2} + 2 L \frac{D}{A^2} + 2K \frac{1}{A}\d(x) \label{4'}
\eea
We stress that some care is required in performing algebraic manipulations
 on these equations, because of the nilpotent nature of many of the quantities
involved.

   Multiplying \reff{2'} and \reff{3'}
by appropriate factors of either $B_+/A$ or $C_-/A$, we find
\bea
\left(\frac{C_-}{A}\right)'\frac{B_+}{A}  &=& 0 \label{2''} \\
\left(\frac{B_+}{A}\right)' \frac{C_-}{A} &=& 0 \label{3''}
\eea
The sum of these two equations can be written as
\beq
\left(\frac{B_+ C_-}{A^2}\right)'  = 0
\eeq
which implies that
\beq
 \frac{B_+ C_-}{A^2} = \g   ~~~, \label{gamma}
\eeq
where $\g$ is a nilpotent constant, so that \reff{1'} becomes
\beq
D -2L  = - \g A \label{1''}
\eeq

  We return now to the fourth of our original set of equations, \reff{4'}.
Rewriting the ratios of the derivatives, and using \reff{2'}, \reff{3'},
and \reff{1''}, we obtain
\beq
AA'' -{A'}^2 = -2 (\g L A + 2 {L}^2) + 2K A \d(x) \label{4''}
\eeq
Following the bosonic example, we let $A= A(|x|)$, and substituting
(where now the prime denotes derivative with respect to the argument of A)
\beq
AA'' + 2 AA' \d(x) -{A'}^2 = -2 \g L A - 4 {L}^2 + 2K A \d(x)  \label{|4''|}
\eeq
 Matching the $\delta$-functions, we get
\bea
A(0)A'(0) &=& KA(0)    \label{5} \\
AA'' -{A'}^2 &=& -2 \g L A - 4 {L}^2  \label{6}
\eea

  To solve \reff{6}, we expand $A$ into its ordinary and nilpotent
parts, $A = f(|x|) + \g g(|x|)$, and solve for $f$ and $g$ separately by
matching coefficients of $\g$.   We find
then at $x=0$
\bea
f(0)f'(0) &=& Kf(0) \\
f(0)g'(0) + g(0)f'(0) &=& Kg(0)
\eea
In order to have a non-trivial solution, we assume that $f(0) \neq 0$,
 and $g(0) \neq 0$, and obtain
\bea
f'(0) = K &,& g'(0) = 0 \label{5'}
\eea
as well as
\bea
ff'' - {f'}^2 &=& -4 {L}^2  \label{7}  \\
fg'' + gf'' - 2 f'g' &=& -2 L f  \label{8}
\eea
Note that \reff{7} is just the bosonic Liouville equation for $f$.

Using standard techniques, we find that there are two sets of solutions
that give physically interesting results.  We detail the first set here
(the second one appears below).
We find for $f$ and $g$
\bea
f(|x|) &=& 2 L c_0\cos(\frac{|x|-c_1}{c_0}) \label{f} \\
g(|x|) &=& 2 L c_0^2 + c_2\sin(\frac{|x|-c_1}{c_0}) +
       c_3 \left[|x| + c_0\cot(\frac{|x|-c_1}{c_0})\right]
        \sin(\frac{|x|-c_1}{c_0}) \non \\
\label{g}
\eea
where the $c_i$'$s$ are arbitrary constants of integration.
   From \reff{5'}, we
determine $c_1 = \frac{1}{2L} \sin^{-1}(-\frac{\k m}{32L})$
 and $c_2= 0$.  We find that $c_3$ is arbitrary.  (Note that choosing
 $c_0 = \frac{1}{2L}$ gives  correspondence
with the bosonic solution.)

    Having determined $A$, we can now find $D$ in terms of $f$.
We find
\beq
D = 2L - \g f \label{D}
\eeq
We turn now to solving for $B$ and $C$.  Using \reff{1'} in \reff{2'} and
\reff{3'}, and
also that $A= f+ \g g$, we have
\bea
\left(\frac{B_+}{f}\right)' &=& 2iL \frac{C_-}{f^2} \label{Bp}\\
\left(\frac{C_-}{f}\right)' &=& -2iL \frac{B_+}{f^2} \label{Cq}
\eea
By using the Liouville solution for $f$, \reff{f}, and defining
 $p = \frac{B_+}{f}$ and $q = \frac{C_-}{f}$, \reff{Bp} and \reff{Cq} become
\bea
p' &=& \frac{i}{c_0} \frac{1}{\cos(\frac{|x|-c_1}{c_0})}q \label{p'} \\
q' &=& -\frac{i}{c_0} \frac{1}{\cos(\frac{|x|-c_1}{c_0})}p \label{q'}
\eea
These can be solved through a change of variables, $P =p + iq$, $Q=p -iq$
where
\beq
\frac{P'}{P} = \frac{1}{c_0\cos(\frac{|x|-c_1}{c_0})} ~~,~~
\frac{Q'}{Q} = -\frac{1}{c_0\cos(\frac{|x|-c_1}{c_0})}
\eeq

By looking separately at $x < 0$ and $x > 0$, we find that the
solutions can be written as
\bea
 P &=& \chi \left[\frac{1 +
\ve(x) \sin(\frac{|x|-c_1}{c_0})}{\cos(\frac{|x|-c_1}{c_0})}\right]  \\
 Q &=& \xi \left[\frac{1 -\ve(x)\sin(\frac{|x|-c_1}{c_0})}
 {\cos(\frac{|x|-c_1}{c_0})}\right]
 \label{Q}
\eea
where $\ve(x) = \Th(x) - \Th(-x)$, $\Th(x)$ is the Heaviside function, and
 $\chi$ and $\xi$ are arbitrary nilpotent constants.  We solve for $B$ and
$C$ from this, defining $\chi + \xi = 2i\a$ and $\chi - \xi =  2\b$, and
find  $\g = \a \b $.

Collecting all the component fields,
 we reconstruct the superfield $T$ as
 $T(x, \th) = A(x) + B_+(x) \th^+ + C_-(x) \th^- + D(x) \th^+ \th^-$, where
\bea
A &=& 2Lc_0\cos(\frac{|x|-c_1}{c_0}) +\g \left[ 2L{c_0}^2
           + c_3|x| \sin(\frac{|x|-c_1}{c_0})
              + c_3c_0\cos(\frac{|x|-c_1}{c_0}) \right]\non \\
\\
B_+ &=& 2Lc_0[ i\a + \b \ve(x)\sin(\frac{|x|-c_1}{c_0})]\\
C_- &=& 2Lc_0[\a \ve(x)\sin(\frac{|x|-c_1}{c_0}) - i \b] \\
D &=&  2L[1 - \g c_0\cos(\frac{|x|-c_1}{c_0})]
\eea
 This can be put in closed form, which we  understand by a Taylor series
expansion, as
\beq
T(x, \th) = 2L(\th^+ - i \b c_0)(\th^- -i \a c_0)
               +  2Lc_0(1+\frac{c_3 \g}{2L} )
              \cos \left[(\frac{|x| - c_1}{c_0}) - \b\th^+ - \a \th^- - \frac{
                       c_3 \g}{2Lc_0} |x| \right]
\label{atrigT}
\eeq

The second solution for $f$ and $g$  is given by
\bea
f(|x|) &=& 2 L c_0\sinh(\frac{|x|-c_1}{c_0}) \label{f2} \\
g(|x|) &=& -2 L c_0^2 + c_2\cosh(\frac{|x|-c_1}{c_0}) +
     c_3\left[ c_0\tanh(\frac{|x|-c_1}{c_0}) - |x|
               \right]\cosh(\frac{|x|-c_1}{c_0})
       \non \label{g2}\\
\eea
and again the $c_i$'$s$ are arbitrary constants of integration.
As before we find  $c_1 = \frac{1}{2L} \cosh^{-1}(\frac{\k m}{32L})$ and
$c_2 = 0$.
 In this case
\bea
 P &=& \chi\left[ \frac{\ve(x) \cosh(\frac{|x|-c_1}{c_0}) -1}
              {\sinh(\frac{|x|-c_1}{c_0})} \right]\\
 Q &=& \xi \left[\frac{\ve(x) \cosh(\frac{|x|-c_1}{c_0}) +1}
              {\sinh(\frac{|x|-c_1}{c_0})} \right]
\eea
with $\chi$, $\xi$ arbitrary nilpotent constants.  We solve for $B$ and
$C$ as above, with $\a$ and $\b$ defined as previously.

Putting it all together, we obtain the second superfield $T$ as
$T(x, \th) = A(x) + B_+(x) \th^+ + C_-(x) \th^- + D(x) \th^+ \th^-$, where
\bea
A &=& 2L c_0\sinh(\frac{|x|-c_1}{c_0})  -\g \left[ 2L{c_0}^2 + c_3 |x|
             \cosh(\frac{|x|-c_1}{c_0})- c_3c_0 \sinh(\frac{|x|-c_1}{c_0})
               \right] \non \\
\\
B_+ &=&  2Lc_0[ i\a \ve(x)\cosh(\frac{|x|-c_1}{c_0}) -\b] \\
C_- &=&  -2Lc_0[\a +i\b \ve(x) \cosh(\frac{|x|-c_1}{c_0}) ]\\
D &=&  2L[ 1- \g c_0\sinh(\frac{|x|-c_1}{c_0})]
\eea
and once again $\g = \a \b$.
 In closed form, this is
\beq
T(x, \th) = 2 L(\th^+ - \a c_0)(\th^- + \b c_0)
              + 2Lc_0 (1 + \frac{c_3 \g}{2L})
                 \sinh \left[ (\frac{|x| - c_1}{c_0})
          + i\a\th^+ -i\b \th^- -\frac{c_3 \g}{2Lc_0} |x| \right]
                      \label{ahypT}
\eeq

\end{document}